\documentclass[sigconf,table]{acmart}

\AtBeginDocument{%
  }

\copyrightyear{2025}
\acmYear{2025}
\setcopyright{rightsretained}
\acmConference[UMAP Adjunct '25]{Adjunct Proceedings of the 33rd ACM Conference on User Modeling, Adaptation and Personalization}{June 16--19, 2025}{New York City, NY, USA}
\acmBooktitle{Adjunct Proceedings of the 33rd ACM Conference on User Modeling, Adaptation and Personalization (UMAP Adjunct '25), June 16--19, 2025, New York City, NY, USA}
\acmDOI{10.1145/3708319.3733675}
\acmISBN{979-8-4007-1399-6/2025/06}

\acmISBN{978-1-4503-XXXX-X/2018/06}

\usepackage{multirow}
\newcommand{\best}[1]{\cellcolor{cyan!25}\textbf{#1}}
\newcommand{\better}[1]{\cellcolor{yellow!25}#1}

\usepackage{framed}            
\definecolor{mdfbg}{gray}{0.93}
\definecolor{mdfline}{gray}{0.80}

\makeatletter
\newenvironment{mdframed}[1][]{%
  \MakeFramed {\advance\hsize -\width \FrameRestore}%
}{%
  \endMakeFramed
}
\makeatother

\begin{document}

\title[Stealthy LLM-Driven Data Poisoning Attacks Against Embedding-Based RAGs]{Stealthy LLM-Driven Data Poisoning Attacks Against Embedding-Based Retrieval-Augmented Recommender Systems}


\author{Fatemeh Nazary}
\authornote{Corresponding author.}
\orcid{0000-0002-6683-9453}
\email{fatemeh.nazary@poliba.it}
\affiliation{%
  \institution{Polytechnic University of Bari}
  \city{Bari}
  \country{Italy}
}

\author{Yashar Deldjoo}
\orcid{0000-0002-6767-358X}
\email{yashar.Deldjoo@poliba.it}
\affiliation{%
  \institution{Polytechnic University of Bari}
  \city{Bari}
  \country{Italy}
}

\author{Tommaso di Noia}
\orcid{0000-0002-0939-5462}
\email{tommaso.dinoia@poliba.it}
\affiliation{%
  \institution{Polytechnic University of Bari}
  \city{Bari}
  \country{Italy}
}

\author{Eugenio di Sciascio}
\orcid{0000-0002-5484-9945}
\email{eugenio.disciascio@poliba.it}
\affiliation{%
  \institution{Polytechnic University of Bari}
  \city{Bari}
  \country{Italy}
}

\renewcommand{\shortauthors}{Nazary et al.}

\begin{abstract}
We present a systematic study of \emph{provider-side} data poisoning in retrieval-augmented recommender systems (RAG-based). By modifying only a small fraction of tokens within item descriptions---for instance, adding emotional keywords or borrowing phrases from semantically related items---an attacker can significantly promote or demote targeted items. We formalize these attacks under token-edit and semantic-similarity constraints, and we examine their effectiveness in both \emph{promotion} (long-tail items) and \emph{demotion} (short-head items) scenarios. Our experiments on MovieLens, using two large language model (LLM) retrieval modules, show that even subtle attacks shift final rankings and item exposures while eluding naive detection. The results underscore the vulnerability of RAG-based pipelines to small-scale metadata rewrites, and emphasize the need for robust textual consistency checks and provenance tracking to thwart stealthy provider-side poisoning. 
\end{abstract}




\maketitle

\section{Introduction}
\label{sec:intro}

\begin{figure*}[!t]
    \centering
    \includegraphics[width=0.65\linewidth]{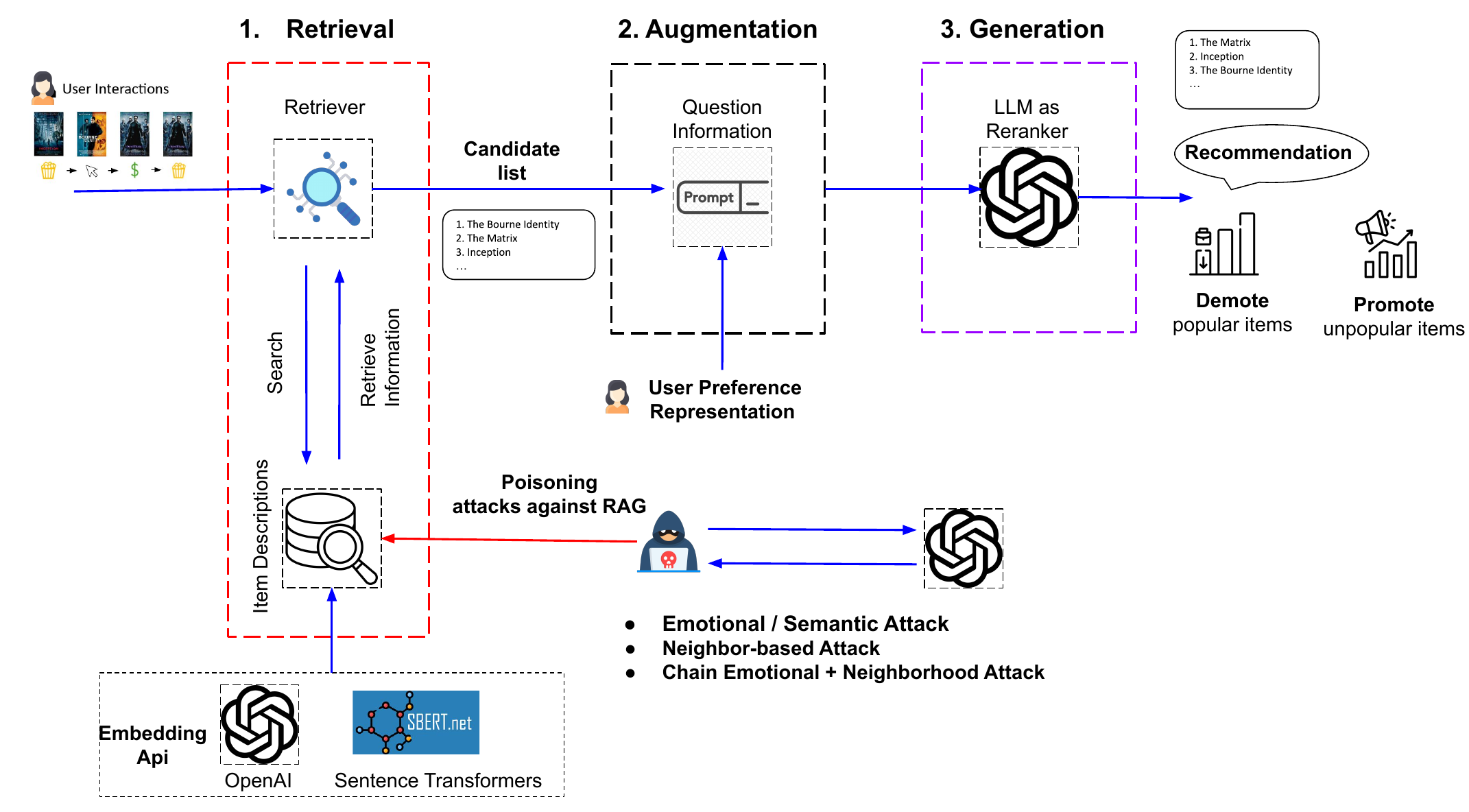}
    \caption{\textbf{High-level RAG architecture in a recommender setting.} 
      A retriever selects candidate items (step~1). 
    An LLM uses these retrieved texts and user queries to re-rank or generate 
    final recommendations (step~2). In our \emph{poisoning} scenario (red arrow), 
    an attacker subtly modifies item descriptions to alter how retrieval 
    and generation perceive items.}
    \label{fig:rag_figure}
\end{figure*}

Retrieval‑augmented generation (RAG) enhances large language models (LLMs) by grounding their outputs in external data sources, such as item reviews or user tags, rather than relying solely on internal parameters~\cite{fan2024survey,deldjoo2024review,deldjoo2024recommendation}. This grounding improves recency and factual accuracy. In fact, industry reports estimate that over 60\% of LLM‑powered search and recommendation systems now use RAG~\cite{fan2024survey,nazary2025poison}. A common RAG-based recommender architecture (Figure~\ref{fig:rag_figure}) retrieves candidate items from an external knowledge store (e.g., a database or corpus of item descriptions) and then uses an LLM to synthesize the retrieved text into final recommendations. While classical methods such as collaborative filtering (CF) can provide a base in the retrieval stage by analyzing user-item interactions, \textbf{embedding-based retrieval} has emerged as a particularly powerful approach in RAG pipelines. Embedding models such as BERT-like encoders or Sentence Transformers can capture an item semantic representation and dynamically decide when to retrieve additional context. By leveraging these embeddings, a recommender system can handle \textit{new} or \textit{infrequently} discussed items, which often appear in long-tail domains. At the same time, these embedding-based retrieval methods offer clear advantages: they provide stronger factual grounding and can adapt to real-time changes in external data. For example, if an item has recently won an award, the RAG pipeline can incorporate that information into the recommendations without retraining the entire model. Notwithstanding their great potential, as more RAG variants adopt embedding-driven approaches, they lean heavily on textual cues, which can open the door to attacks at the data or metadata level. For example, an attacker can subtly \emph{inject} changes into item descriptions (e.g. emotional phrases, negative triggers) to manipulate how both retrieval and generation perceive an item. Unlike classic poisoning that directly tampers with user ratings, \emph{text-based} attacks may remain undetected if they preserve the original semantics. \smallskip


\noindent\textbf{Related Work and Gaps.} LLM vulnerabilities to prompt injection and adversarial prompts have been well documented~\cite{liu2023prompt,wang2024badagent,wei2024jailbroken,chowdhury2024breaking}, however, these studies focus on standalone models rather than recommendation pipelines. In the recommender domain, early poisoning attacks manipulated user ratings or profiles to distort collaborative filtering outputs~\cite{deldjoo2021survey}. More recently, RAG–specific threats emerged: BadRAG~\cite{xue2024badrag} and PoisonedRAG~\cite{zou2024poisonedrag} inject malicious snippets into knowledge bases to warp retrieval or LLM responses. Tag‑based poisoning for RAG recommenders was explored in~\cite{nazary2025poison}, showing that modified user tags can bias item ranking. However, full metadata, such as item descriptions and reviews, remains underexamined. Furthermore, prior methods often rely on simple keyword insertion, which can be detected by basic semantic or stylistic filters. In contrast, we leverage modern LLM rewriting techniques to design coherent, low‑budget alterations that better evade detection~\cite{deldjoo2025toward}.

\vspace{2mm}

\noindent \textbf{Goals and Contributions}
Our primary objective is to formally investigate \emph{provider-side} data poisoning attacks on retrieval-augmented recommenders, focusing on two scenarios: \emph{promotion} (boosting long-tail items) and \emph{demotion} (penalizing highly popular items). In doing so, we introduce the notion of ``textual stealthiness,'' measured through semantic similarity (e.g., SBERT-based) along with overall system-level metrics, to quantify how much an attack can subtly rewrite an item’s description while still achieving malicious ranking shifts. Our main contributions include:
\begin{itemize}
    \item We present a formal definition of provider-side textual rewriting attacks in RAG-based recommendation, framed around two adversarial goals (\emph{promote} vs.\ \emph{demote}).
    \item We use a measurement of ``stealthiness'' by examining how sentence-level semantics change (via SBERT) alongside the overall impact on RS accuracy;
    \item We design and implement \emph{three} distinct attack variations: (i)~\emph{Emotional} edits, (ii)~\emph{Neighbor-based} borrowing, and (iii)~\emph{Chained} rewriting (combining both), all under the same edit-budget constraints.
    \item We empirically evaluate these methods across \emph{two different SoA LLMs} on the  of the MovieLens latest dataset, and demonstrate the potency of small-scale textual manipulations on altering the exposure of carefully selected target items;
    \item We plan to release both \emph{code} and \emph{data} resources to foster reproducibility and future research in adversarial robustness for RAG-based recommender systems.
\end{itemize}


\section{Formal Description of LLM-Driven Data Poisoning Attacks}
The overarching goal of the attacker is to manipulate item visibility within the recommendation pipeline by rewriting the textual descriptions of items. Concretely, we select a subset of items from both the long-tail (unpopular) and short-head (popular) segments, aiming to \emph{promote} the former (i.e., increase their exposure) or \emph{demote} the latter (i.e., decrease their visibility). Formally, for each targeted item $i \in I_{\text{poison}}$ with original description $D_i$, we produce a new description $\widetilde{D}_i$ such that (1)~the token-level change is bounded by $\delta$ (e.g., 10\% of $|D_i|$ tokens may be altered), and (2)~the rewritten text maintains a sufficiently high semantic similarity (e.g., SBERT score above 0.80) to remain stealthy. The attacker’s optimization objective is then to maximize (in the promote case) or minimize (in the demote case) each item’s final ranking position or exposure in top-$k$ recommendations after the system retrains on $\widetilde{D}_i$:
\begin{equation}
\begin{aligned}
    &\max_{\{\widetilde{D}_i\}} \sum_{i \in I_{\text{poison}}} \Delta \Bigl(\text{Exposure}(i)\Bigr) \\
    &\text{subject to} \quad H\bigl(D_i,\widetilde{D}_i\bigr) \le \delta |D_i|, \\
    &\hspace{3.3em} \text{Sim}\bigl(D_i,\widetilde{D}_i\bigr) \ge \sigma_\mathrm{min}.
\end{aligned}
\end{equation}

Here, $\Delta\bigl(.)$ denotes the change in ranking of item~$i$ within the ranking list (either at the retrieval level with top-$N$ or the final recommendation level with top-$k$). In our experiment, we set $N = 50$ and $K = 20$. The function $H(\cdot)$ represents a distance metric, where $\delta$ introduces the notion of \textbf{stealthiness}, ensuring that modifications remain subtle yet effective. We specifically instruct the LLM to modify 10\% of the tokens (token-level distance) while measuring the change at both the \textit{token level},  and the \textit{semantic level} using Sentence-BERT (SBERT)\cite{reimers2019sentence}. In our framework, we designed 3 types of attacks:
\begin{itemize}
    \item \textbf{Emotional Attack:}
    We prompt the LLM to analyze what textual cues make a movie appear ``more popular'' or ``less popular,'' given its original description. The LLM then modifies up to 10\% of the text, injecting emotive or sentiment-laden words (e.g., “exhilarating,” “lackluster”) to steer the item toward promotion or demotion.

    \item \textbf{Neighbor Borrowing:}
    We identify a set of \(n\) \emph{neighbor} items from the opposite popularity segment (i.e., from popular to help promote, or from unpopular to help demote). The LLM then weaves phrases from these neighbors into the target's description, aiming to transfer either highly positive or negative connotations. In our experiments, \(n = 5\).

    \item \textbf{Chain Attack:}
    This approach combines both emotional triggers and neighbor-based snippets in a single pass. While often more powerful, the resulting text can become more visibly altered relative to the original description.
\end{itemize}

\noindent \textbf{Example of Metadata Attack:}

\vspace{5pt}
\noindent
\textbf{Original Description:}

\begin{mdframed}[backgroundcolor=gray!15, linecolor=gray!50]
\emph{``Wings of Hope lifts spirits in trying times. Courage takes flight.''}
\end{mdframed}

\noindent
\textbf{Emotional Attack (Promote):}
\begin{mdframed}[backgroundcolor=cyan!10, linecolor=cyan!50]
``Wings of Hope \textbf{soars} to lift spirits in trying times, where courage takes \textbf{exhilarating} flight. \textit{Uplifting Impact: Embrace the strength within to rise above challenges.}''
\end{mdframed}

\noindent
\textbf{Neighbor Borrowing (Promote):}
\begin{mdframed}[backgroundcolor=yellow!10, linecolor=yellow!50]
``Wings of Hope lifts spirits in trying times. \textit{Experience how Cyber Love blends technology with passion, creating a harmonious blend of optimism and connection.} Courage takes flight.''
\end{mdframed}

\noindent
\textbf{Chain Attack:}
\begin{mdframed}[backgroundcolor=pink!10, linecolor=pink!50]
Merges emotive language (e.g., “exhilarating,” “uplifting”) \emph{and} references to similar ``popular'' neighbors to maximize the embedding shift, e.g.:
``Wings of Hope \textbf{exhilaratingly} showcases resilience \emph{(inspired by Cyber Love's uplifting themes)} to soar beyond challenges.''
\end{mdframed}

\section{Experimental Setup}
\label{sec:system_setup}
We conduct our experiments on top of a retrieval-augmented recommendation (RAG) pipeline that integrates a Large Language Model (LLM) from OpenAI\footnote{Using the official API for text generation and re-ranking} and a Sentence Transformer (ST)\footnote{\url{https://www.sbert.net/}} encoder for embedding-based retrieval. Specifically, the pipeline first employs the Sentence Transformer to select candidate items based on the semantic closeness of item descriptions, and then an OpenAI-based LLM re-ranks these candidates or generates final textual recommendations.

Although we report results primarily on the widely used MovieLens ``ml-latest-small'' dataset in this paper, we have also evaluated the proposed methods on additional benchmark datasets; due to space limitations, we present here only the detailed findings for the MovieLens dataset. 

Within the MovieLens corpus, we categorize items into \emph{long-tail} (unpopular) vs.\ \emph{short-head} (popular) segments, inject adversarial textual edits, and assess the resulting changes in item ranks and system-level metrics (e.g., Recall@\(k\), nDCG@\(k\)). Our system reindexes or retrains on these modified descriptions, thereby simulating real-world scenarios where metadata updates could inadvertently (or maliciously) be incorporated into a live recommender. We build the user profile for the retrieval stage in two ways: (1)~\textbf{Manual} construction using a structured template; and (2)~\textbf{LLM-based} summarization that generates a user’s preferences automatically. In Table~\ref{tbl:promotion_demotion} \textbf{(Tab~2 and~3),} we differentiate these two methods when evaluating final recommendations. All remaining steps in our pipeline (retrieval and re-ranking) remain unchanged.

\section{Results and Discussion}
\label{sec:results}
We now present our key experimental findings, structured around these three main research questions (RQs).

\begin{description}
    \item[\textbf{RQ1:}] Are LLM-based textual attacks effective at pushing a target item ranking up or down (both at retrieval top-$N$ and recommendation stage top-$k$)?
    \item[\textbf{RQ2:}] Does attack efficacy vary across LLMs model e.g., OpenAI vs.\ Sentence Transformer retrieval?
    \item[\textbf{RQ3:}] How do these modifications affect overall recall or nDCG? Can poisoning degrade system-wide performance?
\end{description}

\begin{table*}[!t]
\small
  \centering
  \caption{Side-by-side comparison of \textbf{Promotion} (top) and \textbf{Demotion} (bottom) scenarios for a temporal pipeline. Each scenario lists (A)~Retrieval (temporal), (B)~Recommendation (LLM-based profile), and (C)~Recommendation (Manual profile). Columns show OpenAI vs.\ Sentence Transformer (ST), with Ranking of attacked items (lower = stronger promotion). Bold is the base (no attack), cyan highlights best results, yellow highlights good results.}
  \label{tbl:promotion_demotion}
  \begin{tabular}{@{} ll | ccc | ccc @{}} 
    \toprule
    \multicolumn{2}{c|}{\textbf{Scenario / Pipeline}} 
      & \multicolumn{3}{c|}{\textbf{OpenAI}} 
      & \multicolumn{3}{c}{\textbf{ST}} \\
    \cmidrule(lr){3-5}\cmidrule(lr){6-8}
    & & \textbf{Rank} & \textbf{Recall} & \textbf{nDCG} 
        & \textbf{Rank} & \textbf{Recall} & \textbf{nDCG} \\
    \midrule
    \multicolumn{8}{@{}l}{\textbf{Promotion Scenario}} \\
    \midrule
    \multirow{4}{*}{\shortstack{(A)\\Retrieval}}
      & Original     & \textbf{31.53} & 0.1504 & 0.2101 
                    & \textbf{21.01} & 0.1205 & 0.1615 \\
      & Emotional    & \better{28.65} & 0.1289 & 0.1920 
                    & 33.00         & 0.1191 & 0.1752 \\
      & Neighborhood & \better{28.54} & 0.1486 & 0.2047 
                    & 29.23         & 0.1196 & 0.1698 \\
      & Chain        & \best{25.16}  & 0.1367 & 0.1968 
                    & 32.16         & 0.1241 & 0.1669 \\
    \midrule
    \multirow{4}{*}{\shortstack{(B)\\Rec.\ (LLM)}}
      & Original     & \textbf{7.00}  & 0.0944 & 0.1808 
                    & \textbf{5.27}  & 0.0701 & 0.1561 \\
      & Emotional    & 8.25          & 0.0793 & 0.1838 
                    & 8.00          & 0.0758 & 0.1634 \\
      & Neighborhood & \better{6.67}  & 0.0739 & 0.1546 
                    & 6.40          & 0.0652 & 0.1573 \\
      & Chain        & \best{4.67}   & 0.0830 & 0.1749 
                    & 7.50          & 0.0699 & 0.1548 \\
    \midrule
    \multirow{4}{*}{\shortstack{(C)\\Rec.\ (Manual)}}
      & Original     & \textbf{6.50}  & 0.0853 & 0.1823 
                    & \textbf{4.92}  & 0.0761 & 0.1583 \\
      & Emotional    & \better{6.00}  & 0.0834 & 0.1847 
                    & 8.50          & 0.0749 & 0.1616 \\
      & Neighborhood & \best{3.00}   & 0.0804 & 0.1695 
                    & 7.33          & 0.0661 & 0.1417 \\
      & Chain        & \better{5.89}  & 0.0834 & 0.1743 
                    & \best{1.00}   & 0.0725 & 0.1684 \\
    \midrule
    \multicolumn{8}{@{}l}{\textbf{Demotion Scenario}} \\
    \midrule
    \multirow{4}{*}{\shortstack{(A)\\Retrieval}}
      & Original     & \textbf{25.56} & 0.1504 & 0.2101 
                    & \textbf{26.99} & 0.1205 & 0.1615 \\
      & Emotional    & 22.80         & 0.1299 & 0.1849 
                    & 21.69         & 0.1246 & 0.1679 \\
      & Neighborhood & 24.66         & 0.1414 & 0.1929 
                    & 25.13         & 0.1190 & 0.1674 \\
      & Chain        & 20.60         & 0.1344 & 0.1931 
                    & 25.91         & 0.1251 & 0.1637 \\
    \midrule
    \multirow{4}{*}{\shortstack{(B)\\Rec.\ (LLM)}}
      & Original     & \textbf{5.72}  & 0.0943 & 0.1842 
                    & \textbf{4.95}  & 0.0733 & 0.1543 \\
      & Emotional    & 4.95          & 0.0853 & 0.1966 
                    & 4.40          & 0.0755 & 0.1700 \\
      & Neighborhood & \best{5.87}   & 0.0790 & 0.1759 
                    & 4.75          & 0.0793 & 0.1794 \\
      & Chain        & 5.44          & 0.0747 & 0.1760 
                    & 4.44          & 0.0707 & 0.1599 \\
    \midrule
    \multirow{4}{*}{\shortstack{(C)\\Rec.\ (Manual)}}
      & Original     & \textbf{5.65}  & 0.0972 & 0.1837 
                    & \textbf{4.96}  & 0.0767 & 0.1694 \\
      & Emotional    & 5.53          & 0.0810 & 0.1857 
                    & 4.57          & 0.0670 & 0.1514 \\
      & Neighborhood & 5.63          & 0.0765 & 0.1765 
                    & 4.67          & 0.0817 & 0.1703 \\
      & Chain        & 5.38          & 0.0740 & 0.1748 
                    & \best{5.00}   & 0.0716 & 0.1488 \\
    \bottomrule
  \end{tabular}
\end{table*}

\subsection*{RQ1: Effectiveness of LLM-Based Textual Attacks}
Table~1 (top half) presents results for the \emph{promotion} scenario. The bold “Original” rows provide baseline ranks against which each attack variant (Emotional, Neighborhood, Chain) can be compared. A successful promotion reduces the rank value, indicating an item is placed closer to the top of recommended lists. In several cases, \textsc{Chain} rewriting reduces ranks from approximately 7.0 to around 4.7, whereas \textsc{Emotional} and \textsc{Neighbor} approaches achieve more moderate improvements. These findings validate that even a modest injection of sentiment-laden descriptors or borrowed phrases significantly influences item visibility.

Turning to the \emph{demotion} scenario (Table~1 bottom half), the goal is to push popular items into lower positions (hence, a successful attack increases rank). Chain-based edits again elicit the largest rank changes, demonstrating the capacity of compound strategies—merging emotional cues with neighbor-based snippets—to degrade targeted items more substantially. Thus, we conclude that small-scale textual rewrites are demonstrably effective in shifting final recommendations.

\subsection*{RQ2: Comparison of OpenAI vs.\ Sentence Transformer Retrieval}
An additional observation arises when contrasting the OpenAI columns with the Sentence Transformer (ST) columns. The OpenAI-based pipeline exhibits increased sensitivity to the introduced textual modifications. For instance, in the \emph{promotion} scenario, a change from a rank of 7.0 to approximately 4.7 is relatively large, whereas the corresponding ST scenario occasionally reverses the direction of movement or produces more modest variation. These disparities highlight how generative re-ranking can amplify subtle language cues or signals introduced by adversarial text rewriting. Moreover, the propensity of OpenAI to rely on nuanced phrasing suggests that even brief “trigger” terms can be disproportionately influential, especially compared to a more static embedding architecture.

\subsection*{RQ3: Impact on System-Wide Recall and nDCG}
Beyond item-specific ranking, Table~1 also reports Recall and nDCG. Notably, these global performance metrics do not consistently suffer drastic declines, with the maximum observed drop typically limited to a few percentage points. While such localized attacks primarily disrupt the visibility of a targeted subset, large-scale poisoning—where a significant fraction of items are manipulated—could lead to more pervasive performance deterioration. This aligns with related work demonstrating that simultaneous metadata rewrites on a broader scale can substantially undermine system accuracy~\cite{xue2024badrag}. In short, although the system’s global fidelity remains relatively intact for sparse attacks, the localized impact on individual item positions seems to be more pronounced.

Overall, these results confirm that retrieval-augmented recommender systems are vulnerable to data poisoning via concise textual edits. Small-scale, stealthy manipulations—particularly those that combine emotive triggers and neighbor-based phrasing—can effectuate substantial ranking shifts without severely compromising system-level metrics. The heightened sensitivity of LLM-driven pipelines underscores the importance of developing robust checks on textual provenance and integrity to mitigate provider-side poisoning attempts.

\section{Conclusion}
\label{sec:conclusion}
Our work demonstrates that carefully designed textual perturbations (modifications) in item metadata can strategically alter recommendations in Retrieval-Augmented Generation (RAG) systems, emphasizing the need for robust textual provenance checks. Through a systematic exploration of different attack strategies—including \textit{emotional rewording}, \textit{neighbor-based borrowing}, and \textit{hybrid chaining}—our experiments reveal that even \textbf{small-scale semantic manipulations} can effectively boost the visibility of long-tail items or suppress popular ones, often while remaining stealthy and difficult to detect. These findings underscore the potential provider-side vulnerabilities in RAG-based pipelines and the necessity of defensive measures to safeguard recommendation integrity.

\begin{acks}
The authors acknowledge partial support of the following projects: OVS: Fashion Retail Reloaded and Lutech Digitale 4.0.
\end{acks}


\bibliographystyle{ACM-Reference-Format}
\bibliography{refs}

\end{document}